\documentstyle[12pt,aaspp4,psfig]{article}
\begin{document}

\def\degrees{^\circ }
\def\etal{{\it et al.} }
\def\cf{{\it cf.} }
\def\bv{B-V}
\def\acorr{$\omega(\theta)\ $}
\def\acorrs{$\omega(\theta)$'s\ }
\def\ie{{\it i.e.} }
\def\eg{{\it e.g.} }
\def\spose#1{\hbox to 0pt{#1\hss}}
\def\ltsim{\mathrel{\spose{\lower 3pt\hbox{$\mathchar"218$}}
 \raise 2.0pt\hbox{$\mathchar"13C$}}}

\newbox\grsign \setbox\grsign=\hbox{$>$} \newdimen\grdimen \grdimen=\ht\grsign
\newbox\simlessbox \newbox\simgreatbox
\setbox\simgreatbox=\hbox{\raise.5ex\hbox{$>$}\llap
     {\lower.5ex\hbox{$\sim$}}}\ht1=\grdimen\dp1=0pt
\setbox\simlessbox=\hbox{\raise.5ex\hbox{$<$}\llap
     {\lower.5ex\hbox{$\sim$}}}\ht2=\grdimen\dp2=0pt
\def\gtorder{\mathrel{\copy\simgreatbox}}
\def\ltorder{\mathrel{\copy\simlessbox}}
\def\simgreat{\mathrel{\copy\simgreatbox}}
\def\simless{\mathrel{\copy\simlessbox}}

\title{On the Spatial Distribution of Stellar Populations in the Large
Magellanic Cloud$^1$}

\author{Jason Harris and Dennis Zaritsky}
\affil{UCO/Lick Observatory and Department of Astronomy and
Astrophysics}
\affil{Univ. of California, Santa Cruz, CA, 95064}
\affil{E-Mail:jharris@ucolick.org, dennis@ucolick.org}
\vskip 1in

$^1$ Lick Bulletin No. 1385

\abstract{We measure the angular correlation function of 
stars in a region of the Large Magellanic Cloud (LMC) that
spans 2.0$^\circ \times 1.5^\circ$. We find that the correlation
functions of stellar populations are represented well by exponential
functions of the angular separation for separations between 2 and 40
arcmin (corresponding to $\sim$ 30 pc and 550 pc for an LMC distance
of 50 kpc).  The inner boundary is set by the presence of distinct,
highly correlated structures, which are the more familiar stellar
clusters, and the outer boundary is set by the observed region's size
and the presence of two principal centers of star formation within the
region.  We also find that the normalization and scale length of the
correlation function changes systematically with the mean age of the
stellar population. The existence of positive correlation at large
separations ($\sim 300$ pc), even in the youngest  population, argues
for large-scale hierarchical structure in current star formation. The 
evolution of the angular correlation toward lower normalizations
and longer scale lengths with stellar age argues for the dispersion of
stars with time.  We show that a simple, stochastic, self-propagating
star formation model is qualitatively consistent with this behavior of
the correlation function.}

\keywords{Magellanic Clouds --- stars: evolution --- stars: formation --- stars: kinematics}

\section{Introduction}\label{sec:intro}

The evolution of the stellar component of a galaxy can be separated
into the star formation history, the internal evolution of those
stars, and the dynamical behavior of the system. The study of the star
formation histories of local galaxies, in which individual stars are
resolved, is increasing rapidly in sophistication (cf. \cite{apa97},
\cite{geha98}, \cite{tol98}) and is providing evidence for a wide
variety of histories (see Mateo 1999 for a review). Likewise,
the theory of stellar evolution is one of the triumphs of modern
astrophysics. In contrast, less is known about the initial
distribution and subsequent dynamical evolution of stars within these
systems. Although most stars form
as members of stellar associations (\cite{gom93}, \cite{mass95}),
which are themselves distributed in a way that suggests 
hierarchical structure (\cite{mns53}, \cite{ee96}), the quantitative 
details of the initial stellar distribution and its subsequent evolution
remain only weakly constrained on large scales. Stellar associations,
which are typically unbound (\cite{bla64}, \cite{kon94}) and loosely
defined, will become increasingly dispersed over time and mix with any
true field population, making it difficult to associate older stars
with a localized star formation event.  Even simple issues, such as
the relative number of stars that form in and out of associations,
remain unresolved. 

The detailed study of star forming regions and stellar kinematics has,
by necessity, been confined primarily to star forming complexes in our
galaxy (cf. \cite{gom93}, \cite{lar95}, \cite{bate98}, \cite{gl98}).  
However, external galaxies may provide a contrasting environment, a
viewpoint that is independent of our position within the Milky Way,
and an opportunity to examine the behavior of star formation on
galactic scales. Galactic-scale star formation patterns have been
studied using Cepheids in the Large Magellanic Clouds (\cite{ee96}),
the spatially resolved star formation history in nearby dwarf galaxies
(\cf \cite{dp97}), and the distribution of the most active centers of 
current star formation in more distant galaxies (\cf \cite{hu98}). All
of these studies focus on recent star formation (age $\simless
5\times10^8$ years) and therefore provide more information on the
distribution of current star formation than on the subsequent
dynamical evolution of stellar populations or on any relation that
might exist between past and current star formation. Our goal here is
to extend the study of star clustering over a wider range of stellar
populations, larger areas, and greater lookback times. In particular,
we provide a quantitative description of the distribution of different
stellar populations in the Large Magellanic Cloud (LMC). 

The LMC is an ideal laboratory for this study because it contains a
variety of stellar 
populations, is relatively nearby (\cite{mf98}), has low average
dust extinction (\cite{hzt97}), is close to face-on
(inclined by $\sim30\degrees$ to the line of sight, \cite{wes97}), 
and has a small line-of-sight depth
(\cite{cc86}). Any patterns that
emerge in the stellar distributions 
should be useful in constraining both the initial distribution
and the dynamical evolution of stellar populations.
We do the following with our observations of
the stellar populations in the LMC: 
(1) quantify the clustering properties using the angular
correlation function;  
(2) measure the clustering properties as a function of mean stellar age; 
(3) identify evidence for hierarchical clustering in the young
population over scales smaller than $\sim 550$ pc; 
(4) qualitatively test a model of stochastic, self-propagating star
formation; 
(5) describe the possibility of measuring the true field star fraction; and 
(6) discuss the impact of dynamical evolution on studies of the
stellar populations.

\section{Data and Analysis} \label{sec:data}

The data used in this study come from our $UBVI$ photometric survey of
the Large Magellanic Cloud (LMC) and the Small Magellanic Cloud
(SMC). Zaritsky \etal (1997) present a description of the survey.  
Our final catalog will consist of astrometry ($\sim$ 0.5 arcsec
positional rms error) and photometry ($\sigma_V=0.09$ mag at $V=20$
mag) for $\sim 25$ million stars.  In this paper, we examine the
section of the LMC discussed by Zaritsky \etal, which is centered at
($5.2^h, -67\degrees.4$) and is approximately $2.0\degrees \times
1.5\degrees$ in extent.  The $\bv,V$ Hess diagram for this region of
the LMC is shown in Figure \ref{fig:cmd}.

\subsection{The Stellar Angular Correlation Function} \label{sec:acf}

\begin{figure}[t]
\vspace{5.0in}
\includegraphics{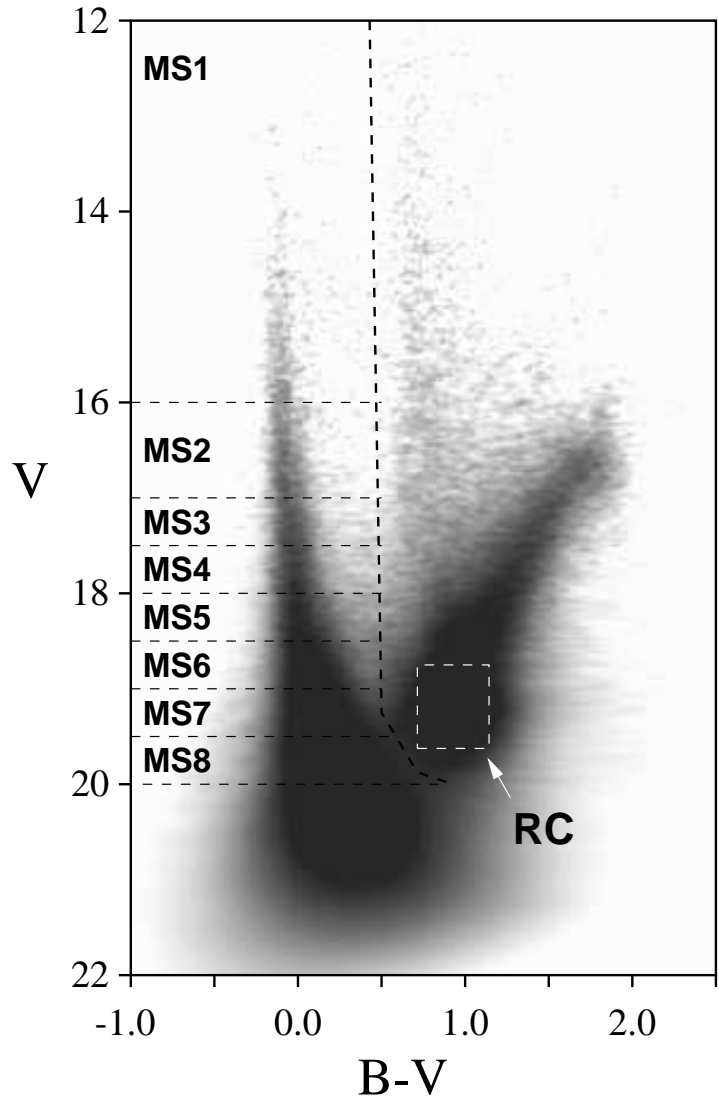}
\caption{A $\bv$, $V$ Hess diagram for our LMC drift scan survey
image.  Over 1.1 million stars are included.  Each star is
represented by a two-dimensional Gaussian with dimensions given by the 
observational uncertainties.  Main sequence stars and red clump stars
have been photometrically isolated as shown.
\label{fig:cmd}}
\end{figure}

A common and natural measurement of the distribution of
objects on the sky is the angular correlation function. 
The evolution of the stellar spatial distribution can be quantified by
examining the behavior of the angular correlation function for
stellar populations of different ages.  We define the angular
correlation function, $\omega(\theta)$, for a given stellar population
to be 

\begin{equation}
\omega(\theta) \equiv \frac{f(\theta)}{f_{R}(\theta)} - 1,
\end{equation}

\noindent
where $f(\theta)$ is the auto-correlation function of the stellar
population, and $f_{R}(\theta)$ is the cross-correlation function
of the stellar population with a randomly distributed population. 
Isolating field stellar populations of a single age in a complex
system like the LMC is nearly impossible (see \cite{ee96} for one
exception) and so we rely on isolating populations of differing {\it
mean} ages.  Assuming that star formation in this region occurred over
at least the past 2 Gyr and had an extended duration (\cite{ber92},
\cite{holt97}, \cite{geha98}), we separate populations that are
sequential in mean age by sorting the upper main sequence (MS) stars
by magnitude. We define upper MS stars to lie in the region of the CMD
bounded by $20.5 - e^{(6.6 - 12(\bv))} < V < 20$ and distribute these 
stars into eight main sequence bins (referred to as MS1 through MS8).
The peculiar definition of the lower $V$ magnitude cut was set
interactively to crudely differentiate MS stars from evolved stars
(\cf Figure \ref{fig:cmd}).  The eight MS bins are defined in Table
\ref{tab:msbin} and labeled in Figure \ref{fig:cmd}. MS1 and MS2 span
a larger range of $V$ magnitudes because there are fewer stars per
magnitude at these luminosities. 

\begin{figure}[t]
\vspace{5.0in}
\includegraphics{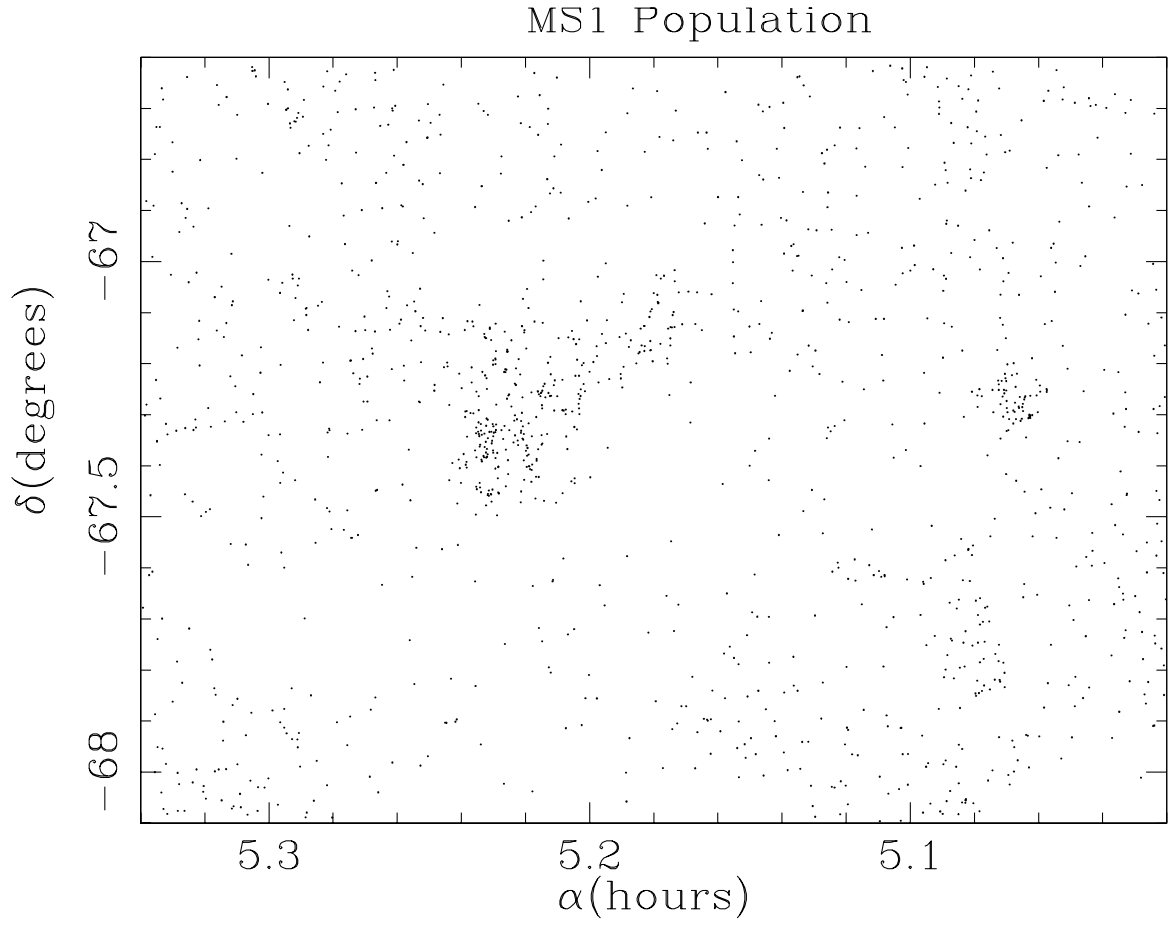}
\caption{The distribution on the sky of the MS1 population, main
sequence stars with $V < 16$ mag (see Table \ref{tab:msbin}).
\label{fig:umscoo}}
\end{figure}

\begin{figure}[t]
\vspace{5.0in}
\includegraphics{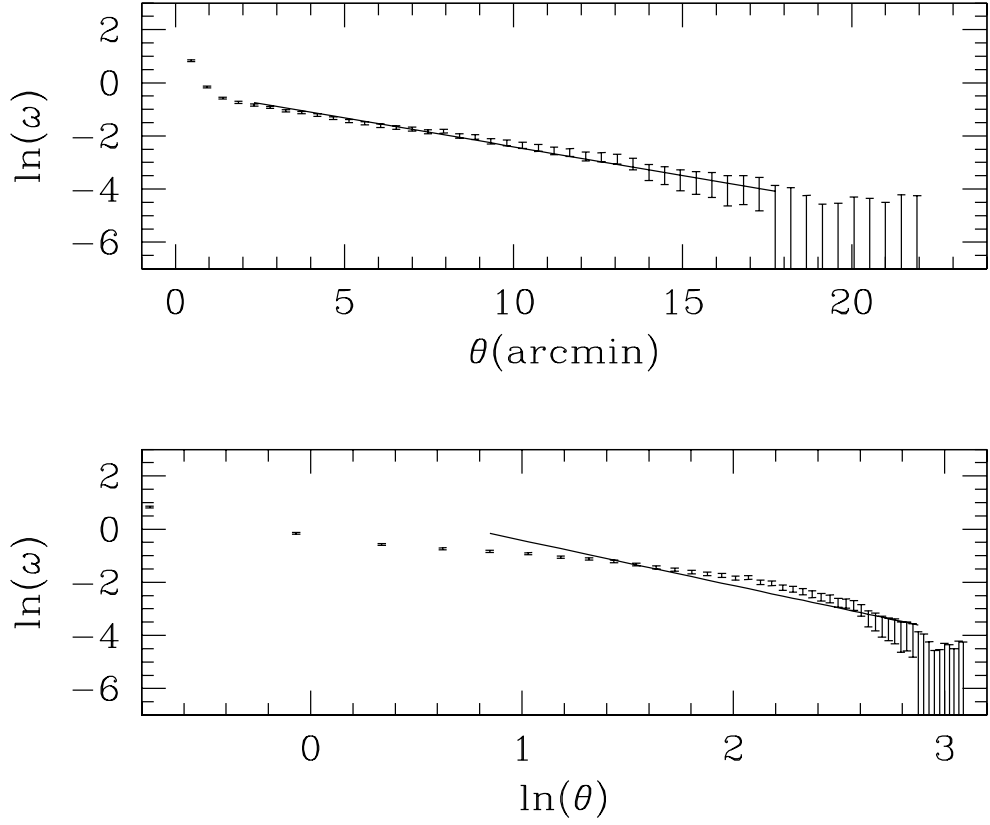}
\caption{The correlation function, $\omega$, of the MS1
population.  Upper panel is $ln(\omega)$ vs. $\theta$.  Lower panel is
$ln(\omega)$ vs. $ln(\theta)$, where $\theta$ is the angular
separation between pairs of stars. 
\label{fig:umscorr}}
\end{figure}

Clustering among the youngest stars (MS1) is evident in Figure
\ref{fig:umscoo}. The quantitative measure of the clustering 
is plotted in Figure \ref{fig:umscorr} in log-linear and log-log
coordinates to illustrate what we find to be a general trend at these
separations: \acorr is better represented by an exponential
function than by a power law. The best-fit exponential and power law
for separations $2 < \theta < 18$ arcmin, which corresponds to
projected physical separations of 30 to 260 pc (adopting $D_{LMC}=50$
kpc, \cite{mf98}), are plotted for comparison. The reduced $\chi^2$ is
1.5 for the exponential fit and 35.9 for the power law fit. This
result is somewhat surprising since clustering of astronomical objects
is often well-described by power law correlation functions (as in
various studies of Galactic star forming regions, \cf \cite{gom93},
\cite{lar95}, \cite{nak98}).  However, our result is not in conflict
with the studies of Galactic associations because those measure the
stellar correlation function on much smaller physical scales
(typically less than 1 pc). For $\theta < 2$ arcmin (30 pc), \acorr
deviates sharply from the exponential fit because of the presence of
tight stellar groupings,  the stellar clusters, and may be better
represented by  a power-law. At large separations, \acorr again
deviates from the fitting function, but \acorr at these separations is
affected by the cross-correlation between the two large stellar
aggregates in this region (\cf Figure \ref{fig:umscoo}) and by the
limited size of the observed region.  

\begin{figure}[t]
\vspace{5.0in}
\includegraphics{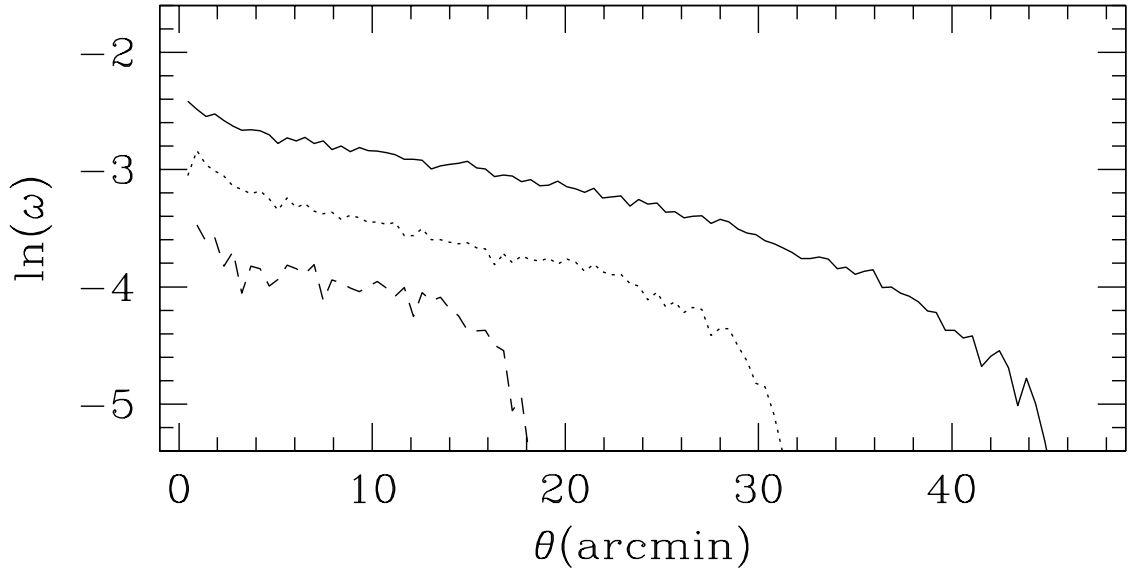}
\caption{Correlation functions of three subregions of the RC
population.  The subregions are concentric with the original observed
region, but their sizes are $1.45\degrees \times 1.45\degrees$ (solid
line),  $1.0\degrees \times 1.0\degrees$ (dotted line), and
$0.5\degrees \times 0.5\degrees$ (dashed line).  The correlation
functions are offset vertically because the correlation is due to a
gradient in the surface density of RC stars.
The cutoff separation at
which the populations become uncorrelated occurs at approximately 0.5
- 0.65 times the angular size of the observed region. \label{fig:edge}}
\end{figure}

To determine the smallest angular separation at which the finite size
of the region begins to significantly degrade the results, we
calculate \acorr for red clump stars ($0.70 < \bv < 1.2$ mag and $18.8
< V < 19.6$ mag (see Figure \ref{fig:cmd})) in three concentric
subregions of sizes  $1.45\degrees \times 1.45\degrees$, $1.0\degrees 
\times 1.0\degrees$ and $0.5\degrees \times 0.5\degrees$. 
We choose these stars for the test because they are distributed nearly
uniformly throughout the region.  Two qualitative oddities are evident
in the measured correlation functions (Figure \ref{fig:edge}).  First,
the \acorr's are systematically offset from each other because there
is a surface density gradient in the red 
clump (RC) population.  The smallest subregion has the least contrast 
between the highest and lowest surface density, so the correlation
values are depressed.  Second, all three correlation functions have a
steep drop at angular separations of 0.5 to 0.65 times the size of the
region.  This drop occurs because of the limited size of the region
and the mathematical construction of the correlation function such
that there is zero correlation at large separations.  We conclude that
if we restrict our study to angular separations that are less than
half the smaller dimension of the observed region (corresponding to
$\theta < 40$ arcmin for our full region), then edge effects will not
affect the shape of the correlation function, even though the
global normalization is undetermined.

\begin{figure}[t]
\vspace{5.0in}
\includegraphics{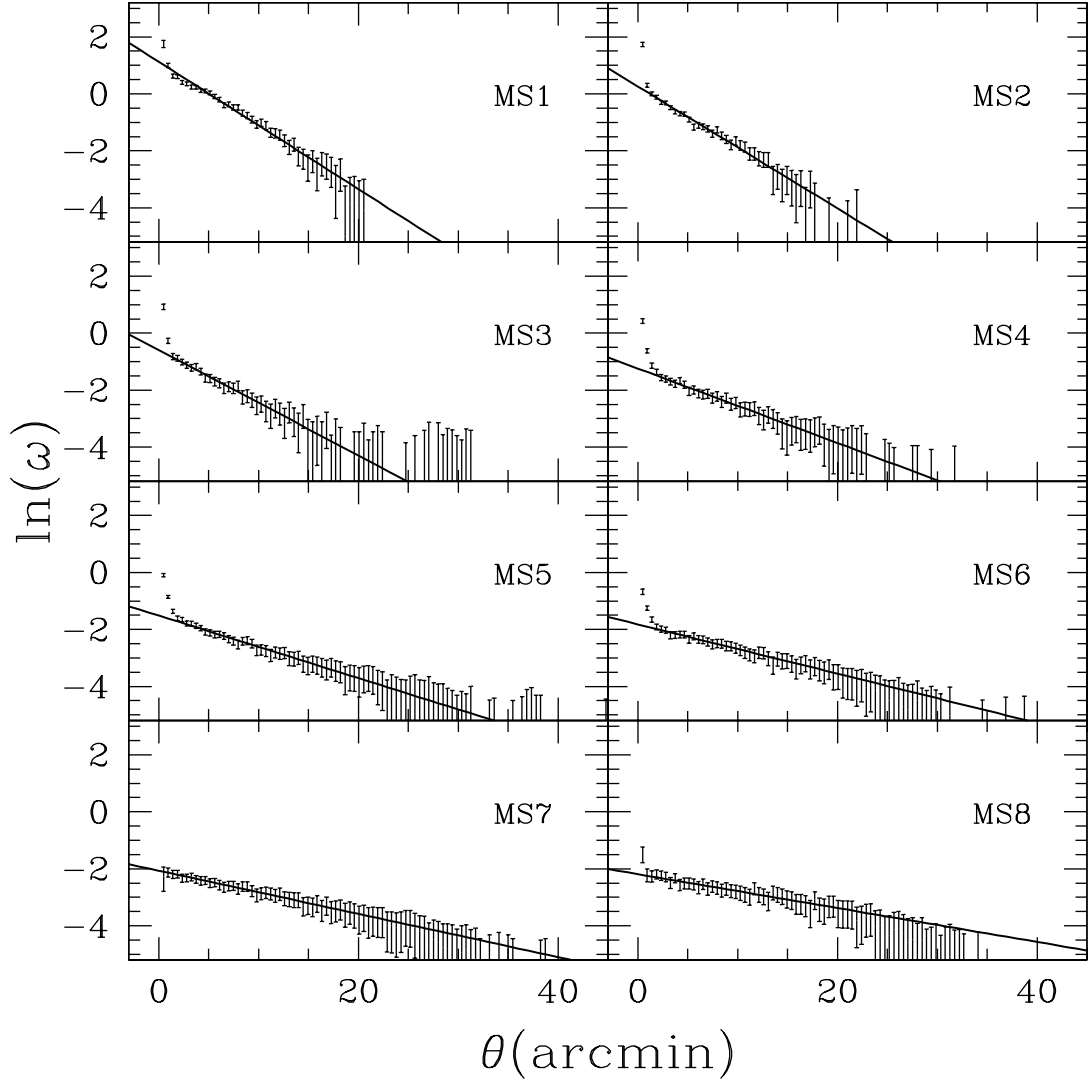}
\caption{Correlation functions of the eight MS populations.  Solid
lines are least squares fits over the range $2 < \theta
< 16$ arcmin for MS1, MS2 and MS3, and $2 < \theta < 20$ arcmin
otherwise. \label{fig:mscorr}}
\end{figure}

\begin{figure}[t]
\vspace{5.0in}
\includegraphics{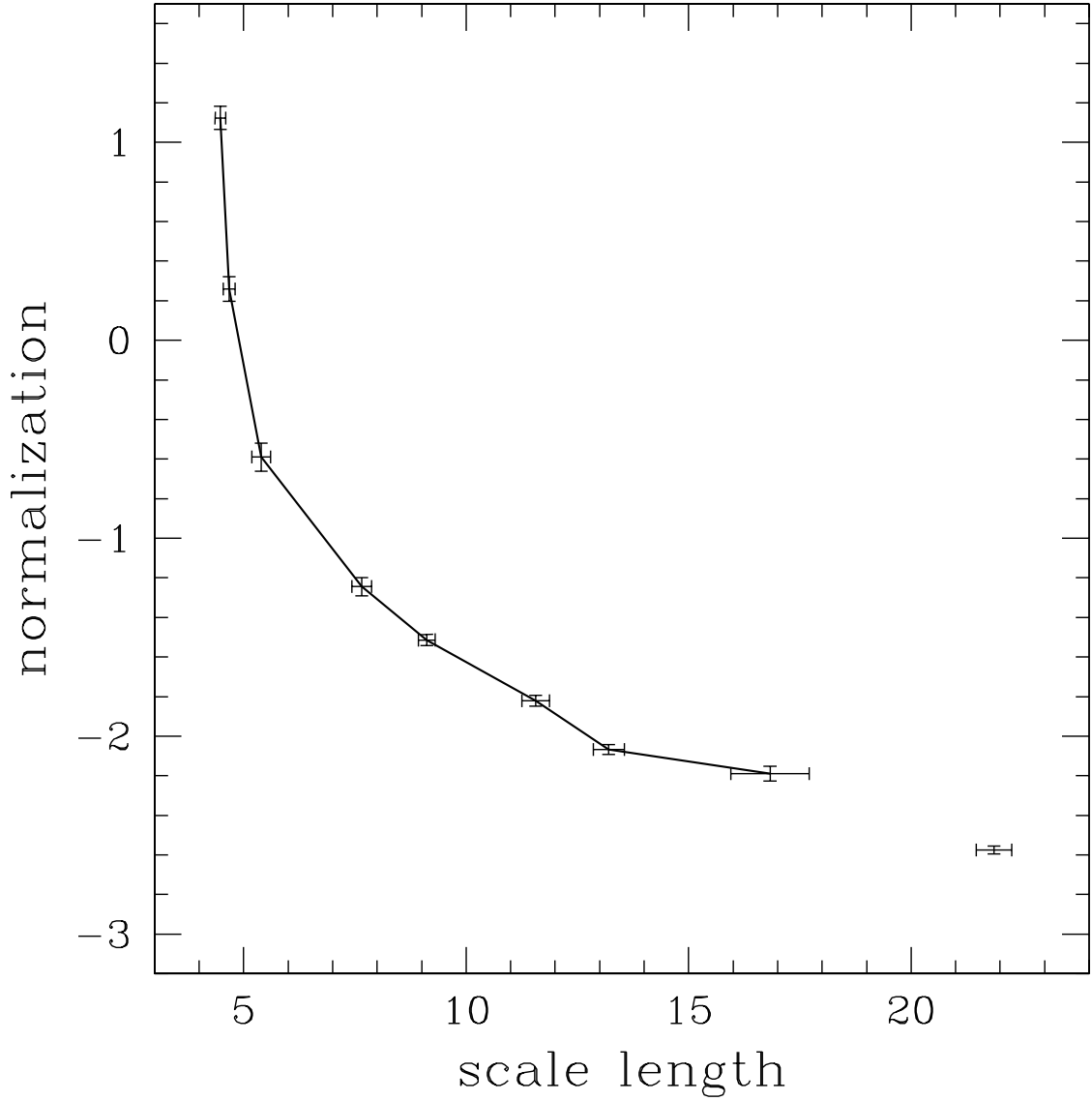}
\caption{Parameterization of the best-fit exponentials to the eight
MS correlation functions plotted in Figure \ref{fig:mscorr}. 
The results for MS1 are in the upper left of the Figure and the
results for MS8 are in the lower right.  The unconnected point is the 
parameterization of the RC population's correlation function. 
\label{fig:msparam}}
\end{figure}

The evolution of clustering is evident in the comparison of \acorr for
the eight MS populations (Figure \ref{fig:mscorr}). We calculate
least-squares fits to each correlation function over the angular
separation range $2 < \theta < 16$ arcmin for MS1, MS2 and MS3, and $2
< \theta < 20$ arcmin for MS4 through MS8.  The $\chi^2$ values for
each fit are $\simless 1$, indicating that the scatter about the
exponential is consistent with the internal uncertainties, and the fits
appear to extrapolate well to larger separations that are $<$ 40 arcmin.
The correlation function scale length increases and its normalization
(which we define as the extrapolation of the best-fit exponential to
zero separation) decreases as the mean luminosity of stars decreases
(Table \ref{tab:msbin} and Figure \ref{fig:msparam}). Because mean age
increases as MS bin magnitude increases, we suggest that the behavior
of the correlation function is the result of the dispersal of stellar
populations as they age (alternatively, the pattern of star formation
may change with time in a way to produce similar behavior).  

\begin{figure}[t]
\vspace{5.0in}
\includegraphics{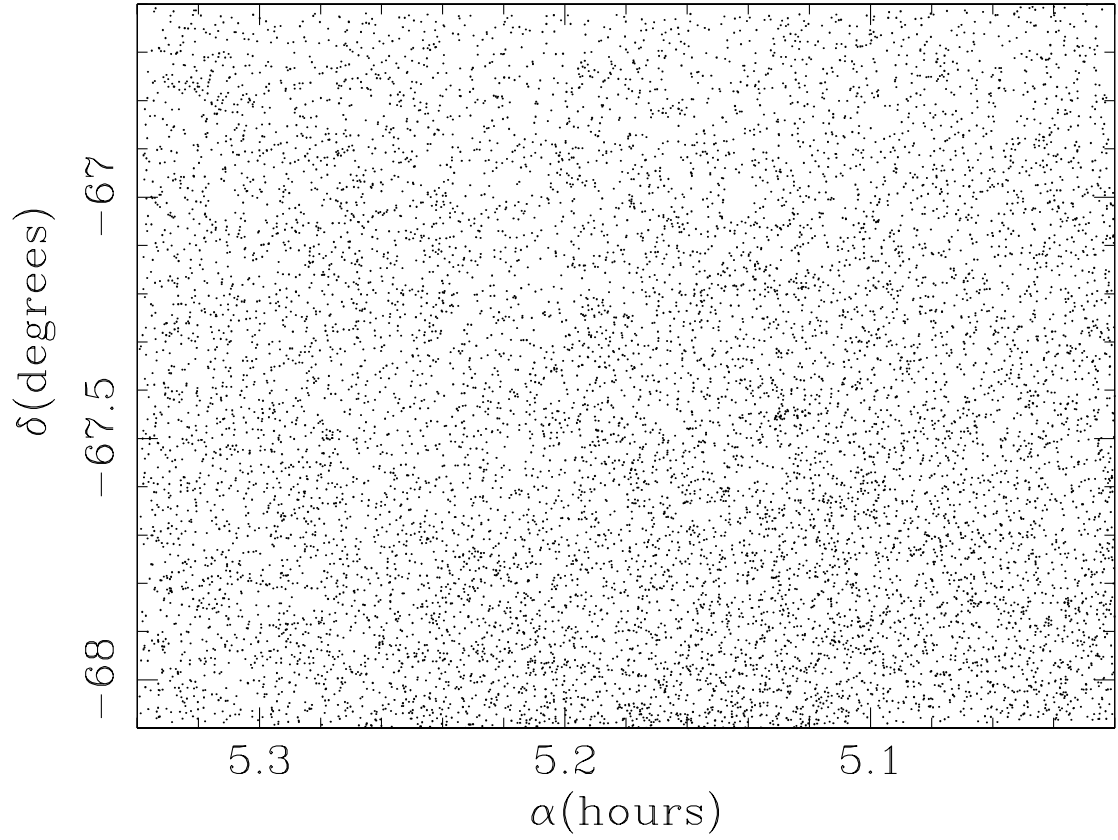}
\caption{The distribution on the sky of the red clump (RC) population.  For
clarity, only 10\% of the $\sim100,000$ RC stars are displayed.
Comparison to Figure \ref{fig:umscoo} illustrates that the MS1 stars are
much more clustered than the RC stars. \label{fig:rccoo}}
\end{figure}

\begin{figure}[t]
\vspace{5.0in}
\includegraphics{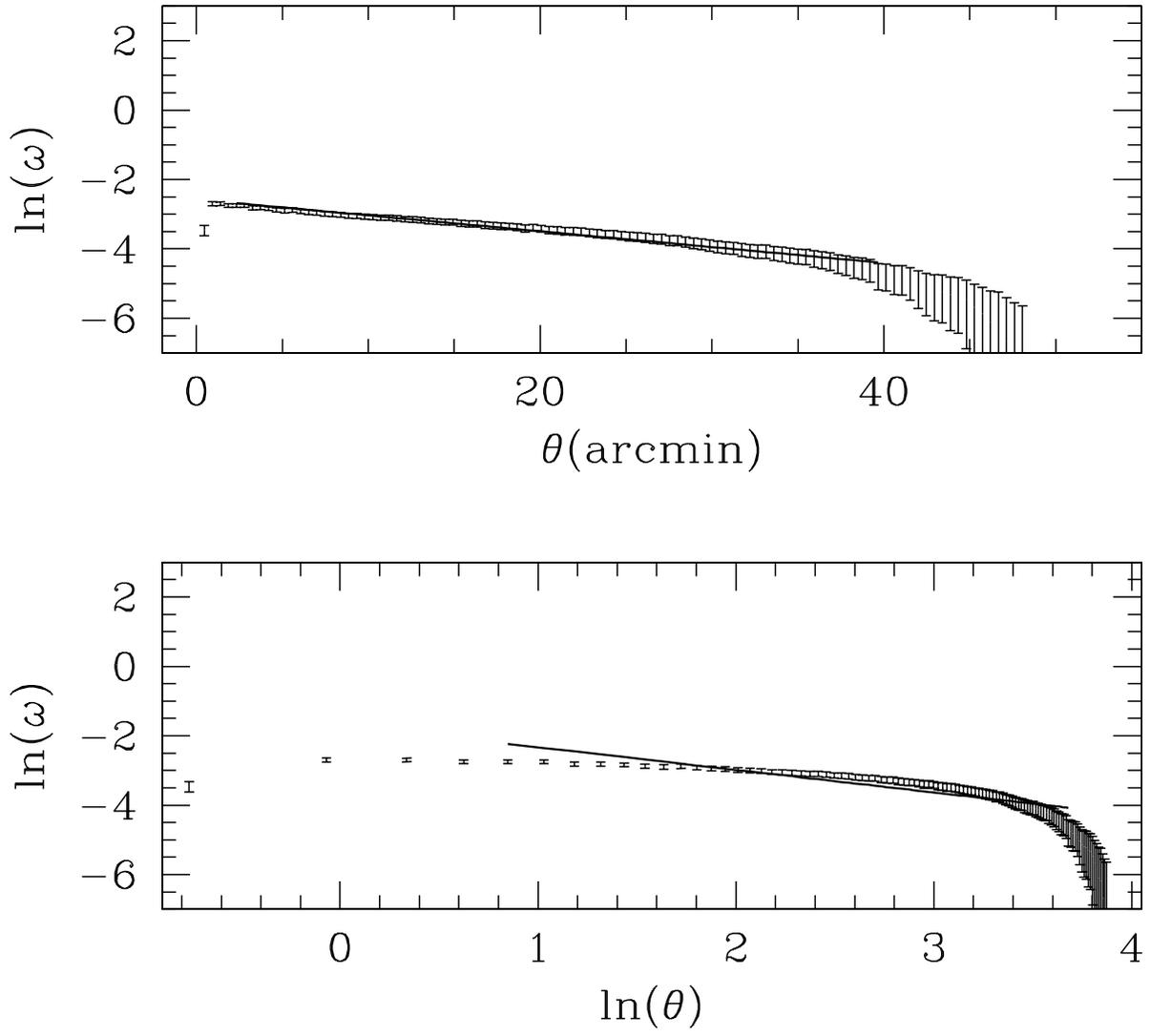}
\caption{The correlation function of the RC population. 
\label{fig:rccorr}}
\end{figure}

To determine whether the correlation function continues to weaken in
even older populations, we calculate \acorr for some of the oldest
stars that we can identify from our data, the red clump
population. The exact age distribution of the RC stars is dependent on
the poorly known star formation history, but the mean age is at least
a few Gyr (for the star formation history derived by \cite{holt97},
73\% of the RC stars are older than 2 Gyr; \cite{cole98}) and
therefore significantly older than the MS populations discussed above.
The spatial distribution of RC stars is shown in Figure
\ref{fig:rccoo} and the corresponding \acorr is shown in Figure
\ref{fig:rccorr}.  The reduced $\chi^2$ for the exponential fit is
0.47 and 6.6 for the power law.  The trend of a decreasing normalization
and a larger scale length with increasing age continues (see
Fig. \ref{fig:msparam}). Finally, the decline in the correlation
function at small radii for the RC stars is most likely an artifact
arising from the greater incompleteness for fainter stars in the dense
cluster cores.  

\section{Discussion}\label{sec:discuss}

We have demonstrated that \acorr for this region of the LMC is well
represented by an exponential function and that it changes
systematically with the mean age of the stellar population.  What do
these observations imply about the evolution of the spatial
distribution of stars in the LMC? Assuming that the clustering
properties of newly formed stars have remained constant over the past
billion years, the systematic change in \acorr implies that dynamical
evolution results in the dispersion of initially highly clustered
stars.  The relatively weaker, but nonzero, \acorr of the RC
population suggests that some clustering signatures may be detectable
for at least a few Gyr.  The quantitative measure presented here
enables us to directly tests models that attempt to explain these
observations.  

In this section, we present highly simplified models that grossly
reproduce the observed behavior. We begin by examining the correlation
function from a stellar population with an exponential surface density
gradient.  We find that such a gradient does not precisely reproduce
the observed correlation among RC stars, a population that is expected
to be well-mixed.  Then we examine simple clustering geometries to 
understand the origin of the observed exponential correlation
function.  We extend this analysis by constructing a model of
self-propagating star formation plus simple dynamical evolution that
reproduces the observed systematic trends in \acorr with stellar
age. While this model is not unique and highly simplified (\eg it does
not include self-gravity or hydrodynamics), it demonstrates that given
plausible assumptions, the observed clustering properties can be
straightforwardly reproduced.  Given the qualitative agreement between
our extremely naive model and the observations, we assert that more
sophisticated models based on the same geneal principles will be able
to quantitatively match the observed behavior.  Finally, we discuss
some implications of our findings on studies of the field star
formation fraction, star formation histories, and stellar dynamics. 

\subsection{A Non-Uniform Underlying Population}\label{sec:rcpop}

\begin{figure}[t]
\vspace{5.0in}
\includegraphics{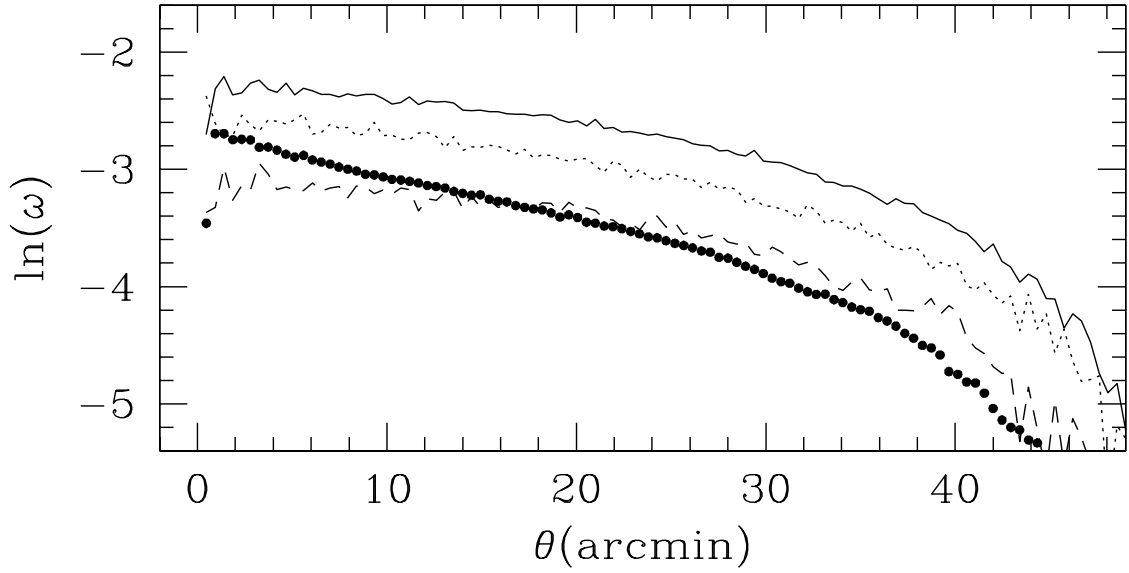}
\caption{Comparison of the RC population correlation function
(points) to that of an artificial exponential-disk population with a
radial scale length of 101 arcminutes (solid line).  We also
constructed artificial populations in which the radial scale length
was 115 arcminutes (dotted line) and 130 arcminutes (dashed line).
\label{fig:gradcorr}}
\end{figure}

To test whether an underlying stellar density gradient can reproduce
the observed correlation function of our most mixed population (the RC
stars), we simulate a stellar population with an exponential surface
density gradient.  We center the artificial population coincident with
the optical center of the LMC ($5.33^h$, $-69\degrees.46$;
\cite{dvf72}), which is about $2\degrees$ south of the observed
region, and adopt a radial length scale of 101 arcmin (Bothun \&
Thompson (1988)). The resulting \acorr of stars placed randomly with
this exponential distribution, and only within a rectangular boundary
corresponding to the observed region has a significantly greater
normalization than that observed (Figure \ref{fig:gradcorr}).
Bothun \& Thompson's radial scale length is based on $B$-band
surface photometry, and so it is possible that the older, redder RC
population has a different radial exponential profile.  Using radial
length scales of 115 arcmin and 130 arcmin, we find that we reduce the
normalization of the correlation function sufficiently but still
cannot recover the observed shape, in particular the rise at smaller
separations ($\theta \simless 16$ arcmin).  We conclude that an
exponential gradient in the surface density of RC stars is
insufficient to entirely explain the observed \acorr.  The residual 
structure in the distribution of RC stars might be the result of
either an RC population that is still dynamically evolving toward a
uniform exponential disk distribution, or of large coherent structures 
in the LMC (\eg spiral arms). This ambiguity should be resolved once 
a larger section of the LMC has been cataloged.

\subsection{The Exponential Correlation Function}\label{sec:exp}

The observed \acorrs are well-fit by exponential functions over a
large range of angular separations.  To determine whether exponential
correlation functions can be generated by extremely simple stellar
distributions of stars, we calculate \acorr for several representative
artificial distributions.  First, we simulate single stellar clusters
with Gaussian and exponential radial density profiles embedded in a  
uniform stellar background. Each simulation contains 6000 stars (a
compromise between spatial resolution and computational time), half of
which are placed according to the cluster profile and half of which
are distributed uniformly across the region. We assign characteristic
radial scales ($\sigma$ for Gaussians, scale length for exponentials)
of 2.5\%, 5\%, 10\% and 25\% of the simulated region's size, to fully
explore the dependence of the correlation function on cluster size
while avoiding edge effects (see \S\ref{sec:acf}).  

\begin{figure}[t]
\vspace{5.0in}
\includegraphics{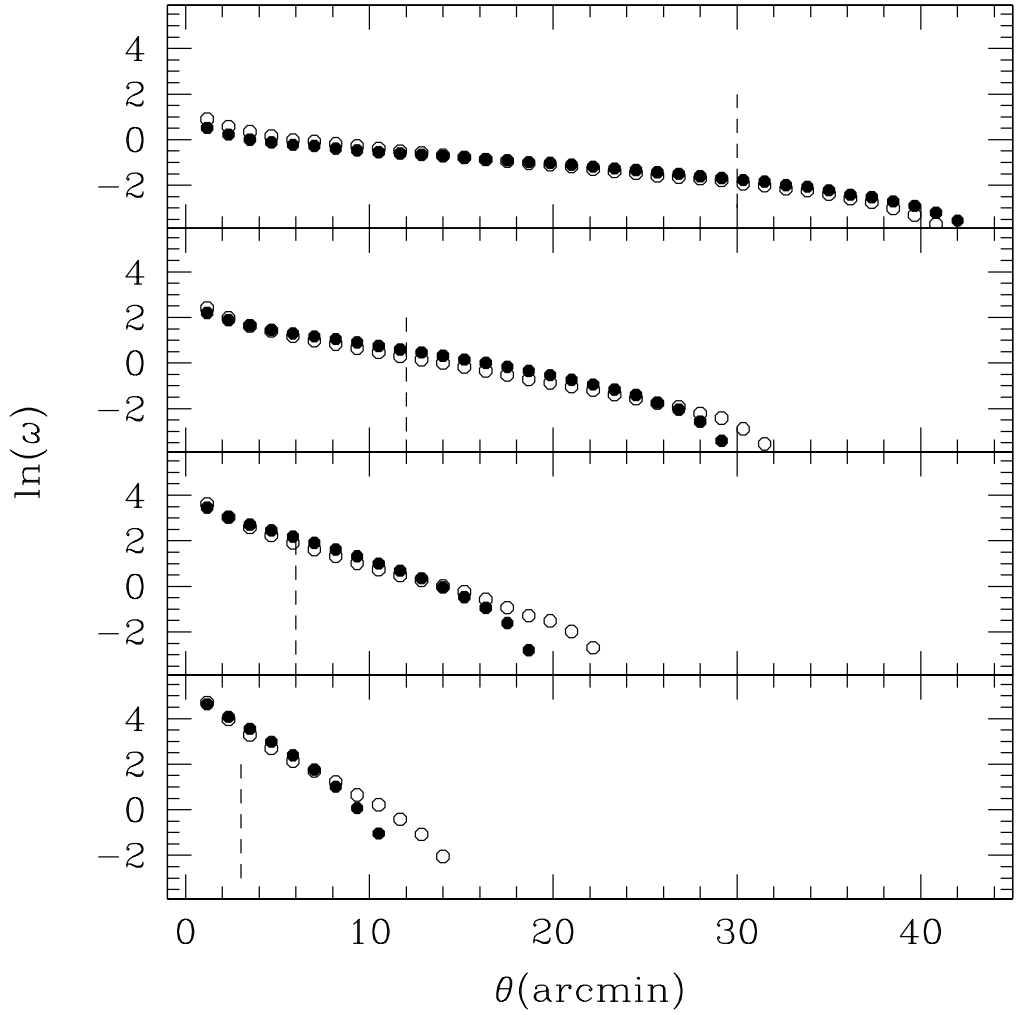}
\caption{The correlation functions of eight simulations containing
individual star clusters.  The filled points represent results
obtained using clusters with a
Gaussian radial stellar density profile, the open points represent
results using exponential radial profiles.  
Each panel contains the results using clusters with 
different characteristic angular size, which is indicated by the dashed
vertical line. \label{fig:clustcorr}}
\end{figure}

The results from these simulations are shown in Figure
\ref{fig:clustcorr}.  The Gaussian and exponential clusters produce
\acorrs that are virtually indistinguishable and consistent with
exponential correlation functions.  Decreasing the characteristic
length scale steepens the correlation function, increases the
normalization, and decreases $\theta_{max}$, the angular separation at
which the stars become uncorrelated ($\theta_{max}$ is a few times the
characteristic length scale in each case). Because observed stellar
clusters in the LMC are typically an arcminute or less in diameter
(\cite{hod88}), the observed positive correlations on scales larger
than a few arcmin must arise from larger single structures or
correlations among clusters.  

\begin{figure}[t]
\vspace{5.0in}
\includegraphics{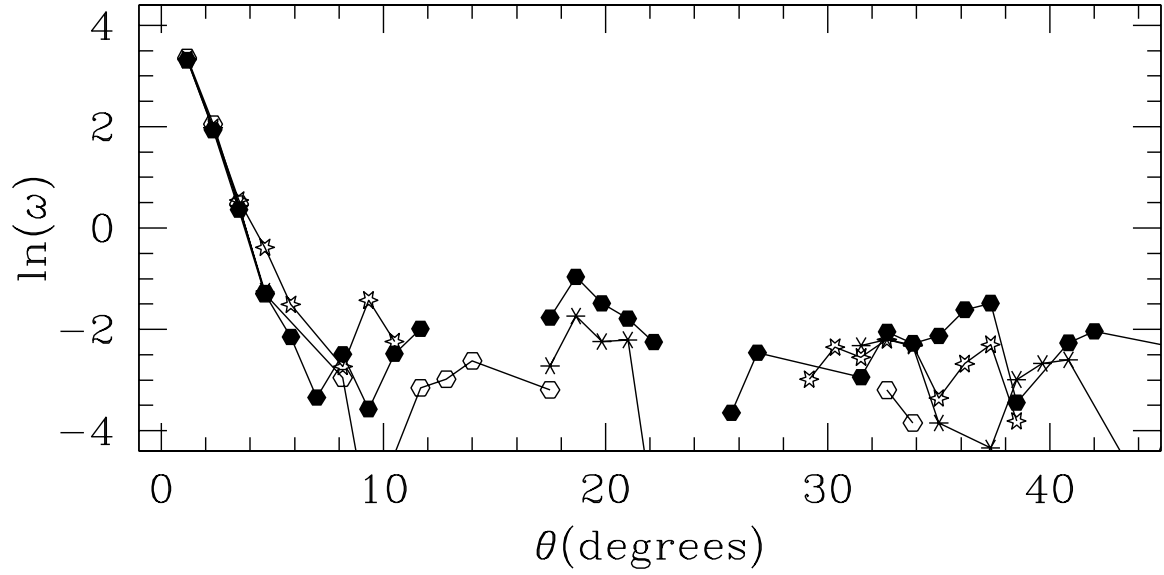}
\caption{The correlation functions of four simulations containing
20 randomly distributed star clusters with exponential radial profiles
and characteristic angular size of 3 arcmin.  The four
simulations differed only in random seed, and each is represented by a
different point style. \label{fig:multicorr}}
\end{figure}

We demonstrate that random distributions of clusters cannot generate
the observed correlation amplitudes at large $\theta$ by simulating 20
randomly placed star clusters, each with an exponential radial profile
and a radial scale length of 1 arcmin.  The resulting \acorrs from
four realizations of this simulation (Figure \ref{fig:multicorr}) are
highly irregular and fail to reproduce the observations. A more
complex model, with correlated cluster positions, is required.

\subsection{The Stochastic Self-Propagating Star Formation Model}

The presence of large scale structure among the young stellar
populations of the LMC is well established (\cite{mns53}, \cite{lh70},
\cite{ee96}).  To investigate whether the simple evolution of such
clustering can reproduce the temporal behavior of the observed
large-scale exponential correlation functions, we use a simple model
of stochastic, self-propagating star formation (SSPSF) to construct
artificial distributions of hierarchically clustered stars.
Traditionally, SSPSF is used to explain the structure of individual
star forming regions (\cite{dop85}, \cite{ftz87}, \cite{ee96}), 
but we employ it here to account for the observed structure of a mixed
population in the field of the LMC.  We allow the stellar
distributions to evolve over time (solely as a result of random
velocities) and test whether such a model reproduces the observed
trend in clustering properties with stellar population age.

Our SSPSF model contains both a field and a cluster mode of star
formation. In both modes, the formation of individual stars is
stochastic. Correlation among clusters will arise if cluster formation
depends on the previous generation of star formation.  There are
various possible physical mechanisms that may produce correlated
cluster formation: stellar winds (\cite{sc99}), a pre-existing
hierarchical structure in the interstellar medium (\cite{lar95}),
gamma-ray bursts (\cite{ef98}), and supernovae explosions
(\cite{tb88}).  In our models, we employ the supernova hypothesis, but
any mechanism that results in correlated clusters over the physical
and temporal scales discussed here will produce similar results. We
trigger cluster formation with supernovae that occur in nearby
clusters.  Our justification for allowing only cluster supernovae to
trigger cluster formation is that field supernovae may not exist in
environments sufficiently dense to form clusters (when field
supernovae are allowed to trigger cluster formation in the models,
the simulated \acorr is weaker than observed).  The goal of our
modeling is not to derive the physical parameters for the SSPSF
process; rather it is to demonstrate that a fairly straightforward
model grounded in processes that are generally believed to occur in
galaxies (dynamical relaxation of stars, correlated star formation,
and self-propagation of star formation) can lead to clustering
properties that are similar to those observed. More specific
inferences will require measurement of the correlation function across
the entire LMC and dynamically self-consistent models that have a more
detailed, physically motivated prescription for star formation (see
\cite{sc99} for one such example). 

The implementation of our model is straightforward.  Stars are formed
at a constant rate, and each star is assigned an initial position, an
initial velocity, and a mass.  A fraction $f$ of stars are formed in
the field, the remainder in clusters. Field star positions are drawn
randomly from an exponential gradient surface density distribution
that matches that observed for the LMC (\cf \cite{bt88}).  Field
stars are assigned zero velocity, in order to preserve the surface
density gradient of their initial distribution (in a more realistic 
model, the gradient would be preserved because the stellar
distribution is in dynamical equilibrium). Cluster stars are
distributed with an exponential radial profile of characteristic
angular scale $\Delta\theta$, centered on one of the simulation's
currently active clusters.  Cluster stars have a Gaussian, isotropic
velocity distribution with mean zero and a one-dimensional velocity
dispersion $\sigma_v$.  The stellar mass is drawn from a Salpeter
initial mass function (IMF) between $2.26 M_{\sun}$ and $100
M_{\sun}$.  The lower mass limit corresponds approximately to the 
faintest stars we study here ($V = 20$ mag), and the upper mass limit
corresponds approximately to the most massive stars observed in the
LMC (\cite{mass95}, \cite{mh98}).  

The stars evolve kinematically in the simplest possible manner. Each
star moves according to its initial assigned velocity until it has
reached the end of its main sequence lifetime, at which time 
the star is removed from the simulation.  The main sequence lifetime
is determined from a quadratic fit to the log($\tau_{MS}$)
vs. log($M$) data tabulated in Fagotto \etal (1994).  Stars that reach
the boundary of the simulated region are reflected back into the
region (this is effectively a periodic boundary condition that
maintains a constant local density).  There are no gravitational
interactions in the simulation. Because OB associations typically have 
mass densities significantly below that at which clusters become
unbound ($\rho\simless0.1 \frac{M_{\sun}}{pc^3}$, \cite{bok34}), 
the stars are not strongly affected by the association's
self gravity, especially
at the angular separations ($> 2$ arcmin) that we are
examining. Finally, we do not include the effect of 
differential rotation in our simulation because the H I rotation curve
of Kim \etal (1998) indicates that our field is within the solid-body
rotation region. Initial models with differential rotation
quickly erased any initial correlations. 
Our survey will eventually include regions of the
LMC for which differential rotation should be important.

Propagating star formation is implemented through 
cluster formation that is triggered by nearby supernovae. If a
cluster star has $M > 8 M_{\sun}$, then that star becomes a
supernova at the end of its lifetime and it may
trigger, with probability $P_t$, the formation of a new cluster within a
spherical shell of inner radius $\theta_t$ and thickness $\delta\theta_t$,
centered on the location of the supernova.
Each viable cluster center exists for
a finite time, $t_{cl}$, after which it ceases to be a
site of localized star formation. The number of active clusters is
self-regulated by having $P_t$ decay
exponentially with the current number of active clusters. 

The physical parameters as described here are summarized in Table
\ref{tab:params}, with some constraints from the literature, where
available.  The parameters that are not constrained by observations
are adjusted so that the simulated \acorrs reasonably match those
observed, although we have not completely explored all of parameter
space.  The simulation is allowed to run for 3 Gyr, which 
encompasses two generations of the longest-lived stars in the MS8
bin ($2.26 M_{\sun}$ corresponds to a main sequence 
lifetime of 1.3 Gyr).  After 3 Gyr, the current projected
positions and the masses of all surviving stars are examined.
The simulated data are analyzed in the same manner as the observations.

\begin{figure}[t]
\vspace{5.0in}
\includegraphics{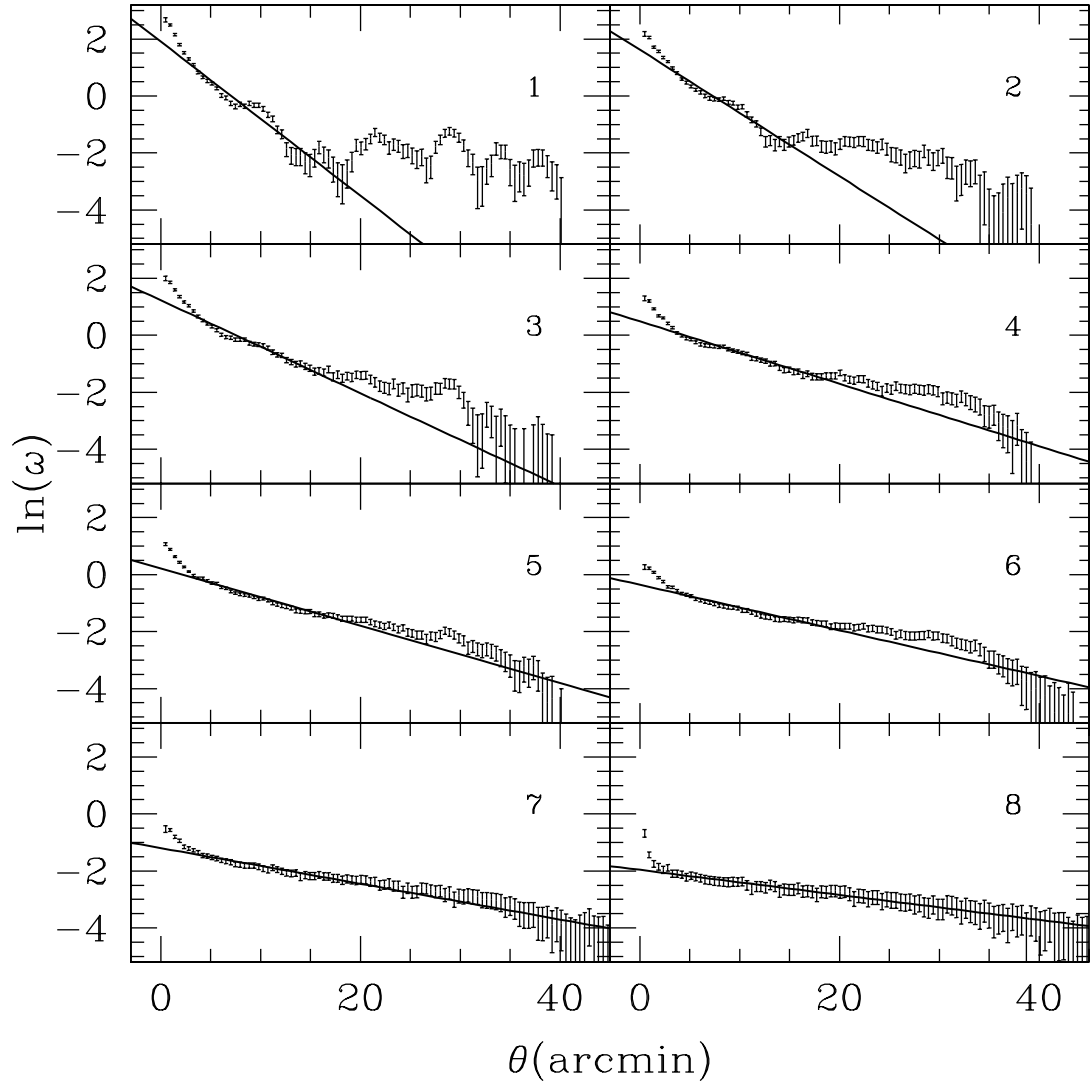}
\caption{The eight MS correlation functions of our best-fitting
SSPSF model. \label{fig:simcorr}}
\end{figure}

\begin{figure}[t]
\vspace{5.0in}
\includegraphics{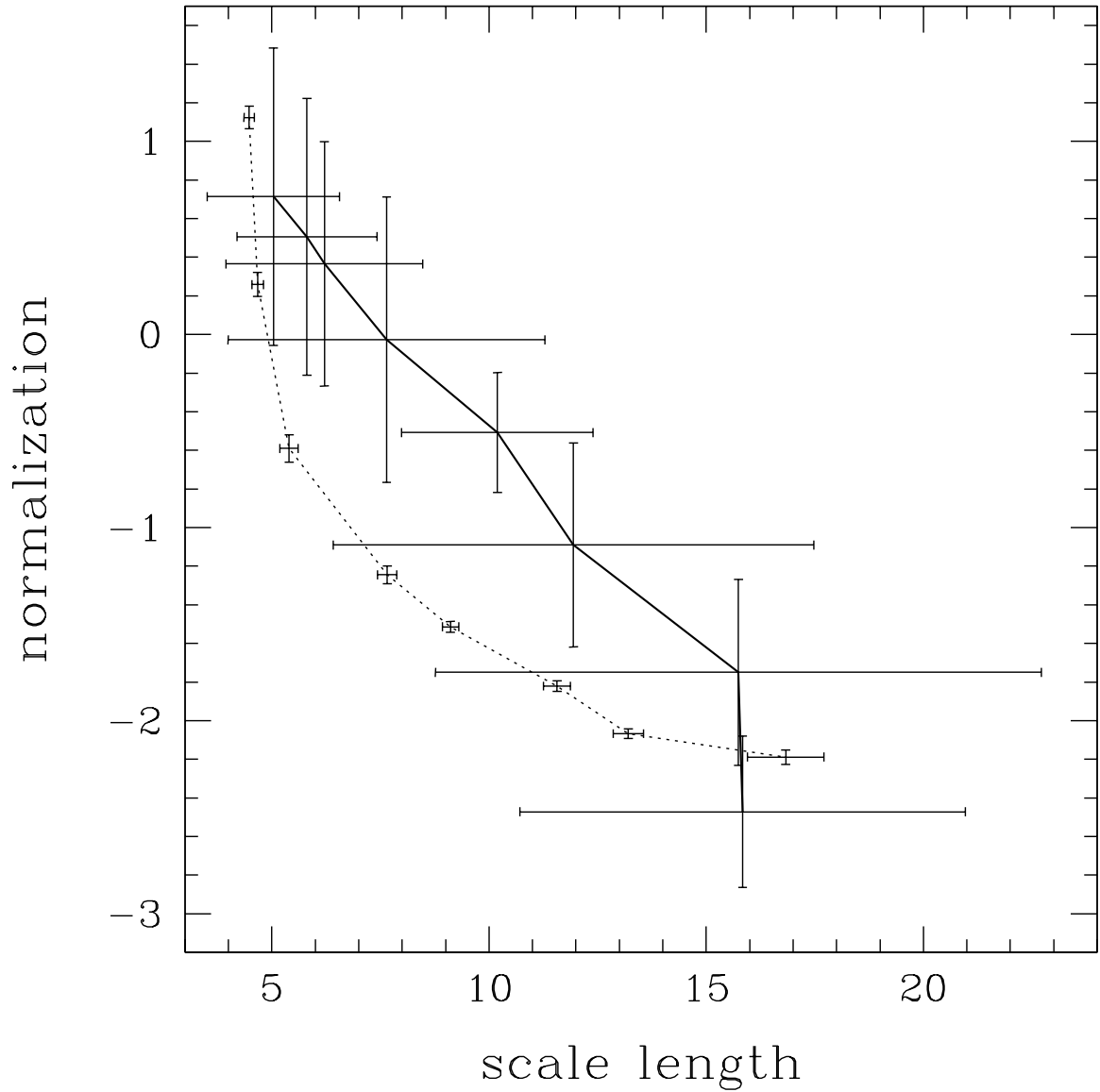}
\caption{Parameterization of the best-fit models to the correlation 
functions of our SSPSF model (solid line).  Also shown is the
parameterization of the observations (dashed line) reproduced from
Figure \ref{fig:msparam} for comparison. 
\label{fig:simparam}}
\end{figure}

The model provides \acorrs that grossly match the observations.
The eight MS correlation functions from one realization of our best
model (\cf Table \ref{tab:params}) are shown in Figure
\ref{fig:simcorr}.  We ran eight realizations of this 
model, varying only the random seed.  Each realization resulted in a
unique sequence of best-fit scale lengths and normalizations, but the
general trend is indicated in Figure \ref{fig:simparam}, in which we
plot the mean values of the fit parameters from our eight
realizations.  The errorbars indicate the standard deviation of the
eight realizations about the mean values.  Also plotted is the
sequence of observed scale lengths and normalizations from Figure
\ref{fig:msparam}.  Although the observed sequence of \acorrs is not
quantitatively consistent with this model, we have reproduced the two
fundamental characteristics of the observed \acorrs:  
(1) the simulated \acorrs are exponential over a large range of
angular separations, and 
(2) the simulated \acorrs decrease in normalization and increase in
scale length for stars drawn from fainter main sequence bins.

More realistic models will need to include:
(1) gravitational interactions among the stars and clusters/associations, 
(2) a global galactic gravitational potential,  
(3) a more realistic star formation history (\cf \cite{alc99}),
(4) a detailed distribution of the interstellar medium and a
prescription for how it determines local star formation, and 
(5) a more detailed and realistic treatment of the self-propagation of
star formation.  

\subsection{Implications} \label{sec:implications}

The current spatial distribution of stars in the LMC is the result of a
mixture of dynamically relaxing and hierarchically clustered stellar
populations of various ages. The distribution is suggestive of an
SSPSF process, although there are alternate ways to produce
hierarchical clustering of stars (e.g., \cite{ee96}, \cite{sc99}).
Our model is sufficiently limited that we have avoided
drawing any but the most general conclusions from it; rather we use
the model to demonstrate one way in which the observed structures may
arise and to examine how such a model may answer additional questions.
Next we describe how 
one might determine the field/cluster
star formation fraction, a quantity that is otherwise very difficult
to measure, and illustrate the limitations of star formation
studies over areas of limited physical scale.

\subsubsection{The Field Star Fraction} \label{sec:fsfrac}

The fraction of stars that form in isolation rather than in clusters and
associations is poorly constrained. The field star formation fraction
in the LMC is claimed to be nonzero because of the existence 
of isolated, massive stars. At their expected velocities of
$\sim 3$ km s$^{-1}$, these stars cannot have traveled from the nearest
cluster or association to their present location during their lifetime
(\cite{mass95}). Massey \etal also found that the logarithmic slope of
the initial mass function is much steeper for this field component
($\Gamma=-4.1\pm0.2$) than for stars in associations and clusters
($\Gamma=-1.3\pm0.3$). This difference in IMF slopes has
tremendous implications for our understanding of galaxy evolution and
star formation, and so it is critical to determine the relative 
numbers of field and cluster stars.
Unfortunately, determining whether stars formed in the field or in
associations becomes intractable on a case-by-case basis for all but the
shortest-lived stars.

Models of the stellar correlation function enable a statistical
approach to the problem.  In our models, the best match to the
observed spatial distribution of stars is achieved when the field star
formation fraction, $f$, is 0.75.  However, by adjusting parameters we
are also able to find an acceptable match with $f=0.5$. Despite the
non-uniqueness of the model, we do find that some field star formation
is necessary to balance the strong correlation of the clusters. This
result may indicate that there is field star formation or that actual
LMC clusters are larger and more irregular than in our
simulations.  One can use similar models to determine how large, or
irregular, associations need to be, or how fast stars need to move, in
order to find models with $f=0$ acceptable.  Although our models are
too preliminary to provide a reliable measurement of $f$, we propose 
that more sophisticated models could utilize \acorr to constrain $f$.  

\subsubsection{Limitations on the Study of the Star Formation History} \label{sec:limit}

The dynamical evolution of stellar populations limits our ability to
recover the star formation histories of local galaxies, both
temporally and spatially. To illustrate, suppose that one observes a
field of angular extent $\theta$ centered on a cluster at a distance
$D$ in which a burst of star formation of duration $\delta \tau$ just
ended.  If stars are dispersing with a typical velocity $v$
km s$^{-1}$, then these stars can propagate from the center of the
cluster to the edge of the observed region in time $t$:

\begin{equation}
t(Myr) = 0.98\Bigl({D \over {\rm pc}}\Bigr)\Bigl({{\rm km/s} \over v}\Bigr)\sin ({\theta / 2}).
\end{equation}

\noindent
If the observed cluster is an OB association in the LMC ($D=50$ kpc),
with typical values $v=3$ km s$^{-1}$ and $\theta=2.4$ arcmin (HST
WFPC2 field), the center-to-edge diffusion time is only 6
Myr. Therefore, stars older than 6 Myr may have dispersed beyond the
observed region and the relative fraction of stars older than 6 Myr is
underestimated.  This bias increases with increasing stellar age.
Without accounting for this effect, one would underestimate the mean
age of the cluster, distort the inferred star formation rate, and
infer either mass segregation or the inward propagation of star
formation. 

The spatial resolution of the star formation histories is also
affected. A stellar population of age $\tau$ 
has typically diffused away from its formation site by an angle
$\theta$: 

\begin{equation}
\theta(\tau)=
1.03 \Bigl({v \over {\rm km/s}}\Bigr) \Bigl({\tau \over {\rm Myr}}\Bigr)
      \Bigl({{\rm pc} \over D}\Bigr),
\end{equation}

\noindent
for small $\theta$.  In the LMC (again assuming a typical $v=3$ km
s$^{-1}$ and $D_{LMC}=50$ kpc), $\theta=0.2\tau$ arcmin.  Resolved
spatial structure in the star formation history on angular scales
smaller than $\theta(\tau)$ is either spurious, or the result of a
structure that is gravitationally maintained (either bound or
sufficiently decelerated by self gravity).  In such a model, our
ability to resolve spatial structure in the star formation 
history is a linear function of the age of the population.

Although these observational biases have been appreciated, 
the observations presented here demonstrate that population
segregation exists out to at least nearly a degree and for stars down
to V = 20 in the LMC. Therefore, all but the studies of the oldest
stars (which one expects to be well mixed) will be affected.

\section{Summary} \label{sec:summary}

We examine the spatial distribution of stars in a region of the LMC
and find evidence for the hierarchical clustering and dynamical
relaxation of stellar populations.  We observe (1) that the stellar
angular correlation functions of different stellar populations are
exponential over separations up to 40 arcmin (the upper limit on the
angular scale of our correlation measurement is set by the current
size of the observed region), and (2) that the parameters of the  
angular correlation function depend on the mean stellar age of the
population (higher normalization and smaller scale length for younger
populations). 

We proceed to interpret these observations through various simple
models. We argue that correlation functions that are exponential over 
separations many times greater than the characteristic cluster size
require correlated cluster positions.  We demonstrate that
hierarchical cluster distributions and the evolution of the
correlation function are consistent with a model of stochastic
self-propagating star formation and the dispersion of stars away 
from the sites of their formation with time. The model has numerous
assumptions and free parameters, so we do not focus on quantitative
results from this modeling.  From the qualitative success of the
model, we conclude that the observed behavior is not unexpected and
that our observations provide detailed quantitative constraints for
more sophisticated models. 

We discuss how these observations may be used to 
determine the fraction of stars that formed in the field rather
than in clusters or associations. Because it has been suggested 
that the field IMF is significantly different than the cluster
IMF (\cite{mass95}), the relative number of stars in the two
populations is critical to reconstructing the star formation history
of the LMC, and presumably of any other galaxy. 
Although our current models are too preliminary to resolve
this issue, more sophisticated models may be able to utilize
the angular correlation functions of stellar populations to reliably
address this issue.

We describe how the observed evolution of the stellar spatial
distribution can affect studies of the star formation history. 
By not accounting properly for the dispersal of stars from the
sites of star formation, one can underestimate the mean age of the
stars formed from a current site of star formation, distort 
the inferred star formation history, and incorrectly infer either mass
segregation or the propagation of star formation. 

In conclusion, the systematic behavior of the angular correlation
functions quantifies what is visually apparent --- that star formation
in the LMC is highly clustered and that older stars are more
diffusely distributed. Nail and Shapely (1953) drew attention to
this fact over 40 years ago, but by quantifying the clustering we
can begin to address additional issues such as the field star
fraction, the importance of propagating star formation, and the effect
of the dispersal of stars with age on other measurements. By combining
the angular correlation function with radial velocity measurements
we will be able determine the age-velocity dispersion relation in a
galaxy other than our own. Eventually, the Magellanic Clouds may
provide less ambiguous, more powerful data than that obtained in our
galaxy with which to understand the dynamical evolution of stellar 
populations within galaxies.

\vskip 1in

\noindent
Acknowledgments:
JH would like to thank Eva Grebel, for her valuable discussions of
issues pertaining to the dynamical evolution of star clusters.
DZ acknowledges financial support from an NSF grant AST-9819576, 
a NASA LTSA grant (NAG-5-3501), fellowships from the David and Lucile
Packard Foundation and the Alfred P. Sloan Foundation, and 
from the Dudley Observatory for the construction of the Great Circle
Camera.

\vfill
\eject

\clearpage

\begin{deluxetable}{c c c c c}
\tablewidth{0pt}
\tablecaption{Main Sequence Bins \label{tab:msbin}}
\tablehead{\colhead{Bin} & \colhead{$V$ range [mag]} & \colhead{Mass
range [$M_{\sun}$]}
& \colhead{Scale Length [arcmin]} & \colhead{Normalization}}
\tablecolumns{5}
\startdata
1 & 10.0 $-$ 16.0 & 30.0 $-$ 11.9 &  4.5 &  1.12\nl
2 & 16.0 $-$ 17.0 & 11.9 $-$ 8.1  &  4.7 &  0.26\nl
3 & 17.0 $-$ 17.5 & 8.1 $-$ 6.5   &  5.4 & $-$0.59\nl
4 & 17.5 $-$ 18.0 & 6.5 $-$ 5.3   &  7.7 & $-$1.24\nl
5 & 18.0 $-$ 18.5 & 5.3 $-$ 4.3   &  9.1 & $-$1.51\nl
6 & 18.5 $-$ 19.0 & 4.3 $-$ 3.4   & 11.6 & $-$1.82\nl
7 & 19.0 $-$ 19.5 & 3.4 $-$ 2.6   & 13.2 & $-$2.07\nl
8 & 19.5 $-$ 20.0 & 2.6 $-$ 2.3   & 16.8 & $-$2.19\nl
\enddata
\end{deluxetable}

\begin{deluxetable}{l r r l l}
\scriptsize
\tablewidth{0pt}
\tablecaption{SSPSF Model Parameters \label{tab:params}}
\tablehead{\colhead{Parameter} & \colhead{Description} &
\colhead{Value} & \colhead{Constraint} & \colhead{Reference}}
\tablecolumns{5}
\startdata
$f$              & fraction of stars in field mode                  & 0.75             & $>0$                     & Massey \etal (1995)       \nl
$n_c$            & Number of clusters at which $p_t$ drops to 0.5   & 0.6              &                          &                           \nl
$\sigma_v$       & line-of-sight velocity dispersion                & 2.75 km s$^{-1}$ & $\simless$ 3 km s$^{-1}$ & Lupton \etal (1989)       \nl 
$\Delta\theta$   & angular size of active star clusters             & 2.3 arcmin       & 0.3 arcmin               & Hodge (1988)              \nl
$\theta_t$       & inner radius of SN trigger annulus               & 0.6 arcmin       &                          &                           \nl
$\delta\theta_t$ & thickness of SN trigger annulus                  & 0.6 arcmin       &                          &                           \nl
$t_{cl}$         & star forming lifetime of clusters                & 1 Myr            & $\sim5$ Myr              & Elmegreen \& Efremov 1996 \nl
\enddata

\end{deluxetable}

\clearpage

\clearpage

\clearpage

\clearpage

\clearpage

\clearpage

\clearpage

\clearpage

\clearpage

\clearpage

\clearpage

\clearpage

\clearpage

\end{document}